\documentclass{ws-ijmpa}

\begin{document}

\markboth{S.N. Yang and S. Kamalov} {Dynamical study of $N
\leftrightarrow \Delta$ transition}

%
\catchline{}{}{}{}{}
%

\title{DYNAMICAL STUDY OF $\gamma^*N \leftrightarrow \Delta$
TRANSITION WITH $N(e,e'p)\pi$ }

\author{\footnotesize SHIN NAN YANG}

\address{Department of Physics, National Taiwan University, Taipei 10617,
Taiwan}

\author{SABIT S. KAMALOV}

\address{Laboratory of Theoretical Physics, JINR Dubna,141980 Moscow
region, Russia}

\maketitle

\pub{Received (Day Month Year)}{Revised (Day Month Year)}

\begin{abstract}
The Dubna-Mainz-Taipei (DMT) dynamical model for pion
electromagnetic production, which describe well the existing data
from threshold up to 1 GeV photon lab energy is presented and used
to analyze recent data in the $\Delta$ region . We find that
within DMT model , the bare $\Delta$ is almost spherical while the
physical $\Delta$ is oblate. The deformation is almost saturated
by the pion cloud effects.

\end{abstract}

 It has been well recognized that the study of the excitations of the
hadrons can shed light on the nonperturbative aspects of QCD. One
case which has recently been under intensive study is the
electromagnetic (e.m.) excitation of the $\Delta(1232)$ resonance.
At low four-momentum transfer squared $Q^2$,  the interest is
motivated by the possibility of observing a $D-$state in the
$\Delta$. The existence of a $D-$state in the $\Delta$ has the
consequence that the $\Delta$ is deformed and the photon can
excite a nucleon through electric $E2$ and Coulomb $C2$
quardrupole transitions, while in a symmetric SU(6) quark model
the electromagnetic excitation of the $\Delta$ could proceed only
via $M1$ transition. In pion electroproduction, $E2$ and $C2$
excitations would give rise to nonvanishing $E_{1+}^{(3/2)}$ and
$S_{1+}^{(3/2)}$ multipole amplitudes. Recent experiments give
nonvanishing ratio $R_{EM} = E_{1+}^{(3/2)}/M_{1+}^{(3/2)} \sim
-0.03$~\cite{Beck97} at $Q^2=0$ which has been widely taken as an
indication of the $\Delta$ deformation.

At sufficiently large $Q^2$, the perturbative QCD (pQCD) is
expected to work. It predicts that only helicity-conserving
amplitudes contribute at high $Q^2$, leading to
 $R_{EM} = E_{1+}^{(3/2)}/M_{1+}^{(3/2)} \rightarrow 1$ and
 $R_{SM} = S_{1+}^{(3/2)}/M_{1+}^{(3/2)} \rightarrow const$.
This behavior in the perturbative domain is very different from
that in the nonperturbative one. It is an intriguing question to
find the region of $Q^2$ which signals the onset of the pQCD.

In the recent measurements \cite{Frolov99,Mertz,Joo}, the e.m.
excitation of the $\Delta$ was studied at $0.1<Q^2<4.0 \,\,GeV^2$
via the reaction $p(e,e'p)\pi^0$. The extracted ratios $R_{EM}$
and $R_{SM}$ remain small and {\it negative}. This indicates that
pQCD is still not applicable in this region of $Q^2$. In this
talk, we want to show that the recent data of Refs.
\cite{Frolov99,Mertz,Joo} can be understood from the dominance of
the pion cloud contribution at low $Q^2$ in the DMT dynamical
model~\cite{KY99} for e.m. production of pion.

The main feature of the dynamical approach to the pion photo- and
electro-production \cite{Yang85} is that the unitarity is built in
by explicitly including the final state $\pi N$ interaction in the
theory, namely, t-matrix is expressed as
\begin{eqnarray}
t_{\gamma\pi}(E)=v_{\gamma\pi}+v_{\gamma\pi}\,g_0(E)\,t_{\pi
N}(E)\,, \label{eq:tgamapi}
\end{eqnarray}
where $v_{\gamma\pi}$ is the transition potential operator for
$\gamma^*N \rightarrow \pi N$ and, $t_{\pi N}$ and $g_0$ denote
the $\pi N$ t-matrix and free propagator, respectively, with $E
\equiv W$ the total energy in the CM frame.

In the (3,3) channel the transition potential $v_{\gamma\pi}$
consists of two terms
\begin{eqnarray}
v_{\gamma\pi}(E)=v_{\gamma\pi}^B + v_{\gamma\pi}^{\Delta}(E)\,,
\label{eq:tranpot}
\end{eqnarray}
where $v_{\gamma\pi}^B$ is the background transition potential
which includes Born terms and vector mesons exchange
contributions, as described in Ref. \cite{UIM99}. The second term
in Eq. (\ref{eq:tranpot}) corresponds to the contribution of a
bare $\Delta$.

We decompose Eq. (\ref{eq:tgamapi}) in the following way,
\begin{eqnarray}
t_{\gamma\pi}(E)=t_{\gamma\pi}^B + t_{\gamma\pi}^{\Delta}(E)\,,
\label{eq:ddecomp}
\end{eqnarray}
where
\begin{eqnarray}
t_{\gamma\pi}^B(E)=v_{\gamma\pi}^B+v_{\gamma\pi}^B\,g_0(E)\,t_{\pi
N}(E)\,,\qquad
t_{\gamma\pi}^\Delta(E)=v_{\gamma\pi}^\Delta+v_{\gamma\pi}^\Delta\,g_0(E)\,t_{\pi
N}(E)\,. \label{eq:decomp}
\end{eqnarray}
The advantage of such a decomposition is that all the processes
which start with the electromagnetic excitation of the bare
$\Delta$  are summed up in $t_{\gamma\pi}^\Delta$.

We evaluate $t_{\gamma\pi}^B$ with pion scattering matrix $t_{\pi
N}$ obtained in a meson-exchange model \cite{Hung94}.
 Note that to make principal value integration associated with
$v_{\gamma\pi}^B$ convergent, we introduce a  dipole form factor
with  cut-off parameter $\Lambda$=440 MeV. The gauge invariance,
violated due to the  off-shell rescattering effects, is restored
by the substitution $J_{\mu}^B \rightarrow J_{\mu}^B - k_{\mu}
k\cdot J^B/k^2$, where $J_{\mu}^B$ is the electromagnetic current
corresponding to the background contribution $v_{\gamma\pi}^B$ and
$k$ is the photon momentum.

For the the $\Delta$ resonance contribution
$t_{\gamma\pi}^\Delta$, the corresponding multipole amplitudes
$A^{\Delta}$ is evaluated using a Breit-Wigner form, as was done
in the isobar model of Ref. \cite{UIM99},
\begin{equation}
A^{\Delta}(W,Q^2)\,=\,{\bar{\cal A}}^{\Delta}(Q^2)\, \frac{
f_{\gamma \Delta}\,\Gamma_{\Delta}\,M_{\Delta}\,f_{\pi \Delta} }
 {M_{\Delta}^2-W^2-iM_{\Delta}\Gamma_{\Delta}}\,e^{i\phi}\,,
\label{eq:BW}
\end{equation}
where $f_{\pi \Delta}(W)$ is the usual Breit-Wigner factor
describing the decay of the $\Delta$ resonance with total width
$\Gamma_{\Delta}(W)$ and physical mass $M_{\Delta}$=1232 MeV. The
expressions for $f_{\gamma \Delta}, \, f_{\pi \Delta}$ and
$\Gamma_{\Delta}$ are taken from Ref. \cite{UIM99}. The phase
$\phi(W)$ in Eq. (\ref{eq:BW}) is to adjust the phase of
$A^{\Delta}$ to be equal to the corresponding pion-nucleon
scattering phase $\delta^{(33)}$. Note that at the resonance
energy, $\phi(M_{\Delta})=0$.

The main parameters in the bare $\gamma^* N \Delta$ vertex are the
${\bar{\cal A}}^{\Delta}$'s in Eq. (\ref{eq:BW}). For the magnetic
dipole ${\bar{\cal M}}^{\Delta}$ and electric quadrupole
${\bar{\cal E}}^{\Delta}$ transitions they are related to the
conventional electromagnetic helicity amplitudes $A^\Delta_{1/2}$
and $A^\Delta_{3/2}$ by ${\bar{\cal
M}^\Delta}(Q^2)=-\frac{1}{2}(A^\Delta_{1/2} + \sqrt{3}
A^\Delta_{3/2})$ and ${\bar{\cal
E}^\Delta}(Q^2)=\frac{1}{2}(-A^\Delta_{1/2} + \frac{1}{\sqrt{3}}
A^\Delta_{3/2}).$
%
%

At the photon point $Q^2=0$,  the bare amplitudes ${\bar{\cal
M}}^\Delta (0)$ and ${\bar{\cal E}}^\Delta (0)$ of Eq.
(\ref{eq:BW}) were extracted from the  best fit to the results of
the recent analyses of Mainz~\cite{HDT} and VPI group~\cite{VPI97}
as shown in Fig. 1 by solid curves. The dashed curves denote the
contribution from $t_{\gamma\pi}^B$ only. The dotted curves
represented the K-matrix approximation to $t_{\gamma\pi}^B$,
namely, without the principal value integral term included. The
numerical values  for ${\bar{\cal M}}^\Delta$ and ${\bar{\cal
E}}^\Delta$, at $Q^2 =0$, are given in Table. 1. Here we also give
"dressed" values obtained using K-matrix approximation. One
notices that the above determined bare values for the helicity
amplitudes, which amount to only about $60\%$ of the corresponding
dressed values, are close to the predictions of the constituent
quark model (CQM). The large reduction of the helicity amplitudes
from the dressed to the bares ones results from the fact that the
principal value integral part of  $t_{\gamma \pi}^B$, which
represents the effects of the off-shell pion rescattering,
contributes approximately for half of the $M_{1+}$ as indicated by
the dashed curves in Fig. 1.

The $Q^2$ dependence of the ${\bar{\cal A}}^{\Delta}$'s is
parametrized by some functional form with parameters determined
from fits to the data of Refs. \cite{Frolov99,Mertz,Joo}. The
resulting model is found to give an excellent description on the
data of Ref. \cite{Frolov99,Mertz,Joo} as well as the e.m. $\pi^0$
production near threshold \cite{Kamalov01}.

\begin{figure}[tbp]
\begin{center}
\epsfig{file=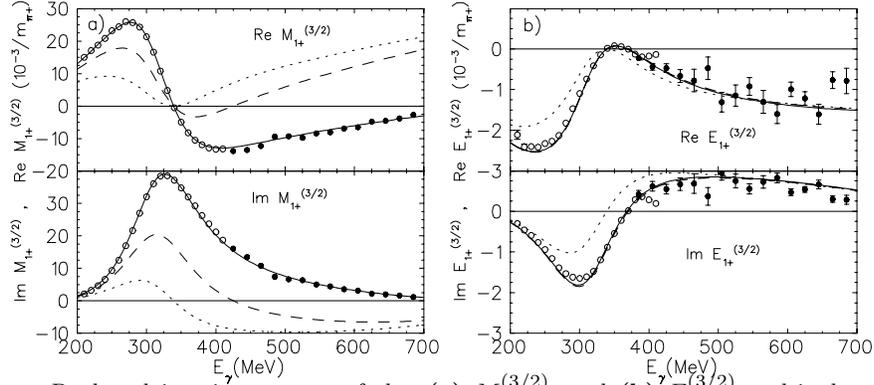,width=2in,height=4.5in, angle=90}
\end{center}
\vspace{-0.6cm} \caption{\small Real and imaginary parts of the,
{\bf (a)} $M_{1+}^{(3/2)}$, and {\bf (b)} $E_{1+}^{(3/2)}$
multipoles. The open and full circles are the results from the
Mainz dispersion analysis~\protect\cite{HDT} and from the VPI
analysis~\protect\cite{VPI97}. Notations for the curves are given
in the text.} \label{panic1}
\end{figure}
\begin{table}[htbp]
\caption {\small Comparison of the "bare" and "dressed" values for
the amplitudes ${\bar{\cal A}}^{\Delta},\, A_{1/2}^{\Delta}$ and
$A_{3/2}^{\Delta}$ (in $10^{-3}\,GeV^{-1/2}$).}
\renewcommand{\tabcolsep}{1.5pc} 
\begin{tabular}{|c|ccc|}
\hline
             Amplitudes            &  "bare"         & "dressed"    & PDG \\
 \hline
 ${\bar{\cal M}}^{\Delta}$  & $ 158\pm 2 $    & $289\pm 2 $  & $293\pm 8$ \\
        ${\bar{\cal E}}^{\Delta}$  & $ 0.4\pm 0.3 $  & $-7\pm 0.4 $ & $-4.5 \pm 4.2 $ \\
\hline
\end{tabular}
\end{table}

Our extracted values for $R_{EM}$ and $R_{SM}$ and a comparison
with the recent results of Refs.~\cite{Frolov99,Mertz,Joo} are
shown in Fig. 2. Our values of $R_{EM}$ show a clear tendency to
cross zero and change sign as $Q^2$ increases. This is in contrast
with the results obtained in Refs.\cite{Frolov99,Sato} which
concluded that $R_{EM}$ would stay negative
\begin{figure}[h]
\begin{center}
\epsfig{file=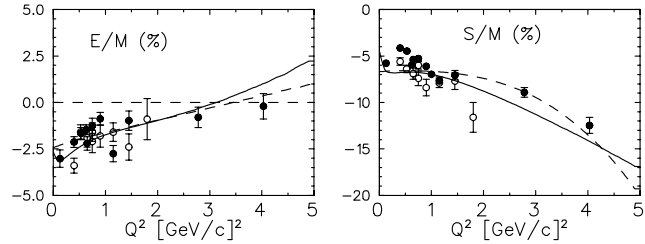, angle=90, width=3.3in}
\end{center}
\caption{ The $Q^2$ dependence of the ratios $R_{EM}^{(p\pi^0)}$
and $R_{SM}^{(p\pi^0)}$ at $W=1232$ MeV. The solid and dashed
curves are the DMT and MAID results, respectively.  Results of our
analysis obtained using MAID and experimental data from
Refs.\protect\cite{Frolov99,Mertz,Joo} ($\bullet$). Open circles
are from Ref.\protect\cite{Joo}.} \label{fig2}
\end{figure}
\noindent with increasing $Q^2$. Furthermore, we find that the
absolute value of $R_{SM}$ is strongly increasing.

In summary, we calculate the $Q^2$ dependence of the ratios
$E_{1+}/M_{1+}$ and $S_{1+}/M_{1+}$ in the
$\gamma^*N\rightarrow\Delta$ transition, with the use of a
dynamical model for electromagnetic production of pion. We find
that the ratio $E_{1+}/M_{1+}$ change the sign at $Q^2 \ge
3.0\,\,GeV^2$. Our results agree well with the recent measurements
from Refs. \cite{Frolov99,Mertz,Joo}, but deviate strongly from
the predictions of pQCD. Our results indicate that the bare
$\Delta$ is almost spherical and hence very difficult to be
directly excited via electric E2 and Coulomb C2 quardrupole
excitations. The experimentally observed $E_{1+}^{(3/2)}$ and
$S_{1+}^{(3/2)}$ multipoles are, to a very large extent, saturated
by the contribution from pion cloud, i.e., pion rescattering
effects.

\end{document}